\documentclass{ws-mpla}
\usepackage[super]{cite}
\usepackage{graphicx}
\usepackage{bm}
\usepackage{url}
\usepackage{caption2}
\usepackage[hidelinks]{hyperref}

\begin{document}

\markboth{E.T. Kipreos \& R.S. Balachandran}
{Assessment of the relativistic rotational transformations}

\catchline{}{}{}{}{}

\title{Assessment of the relativistic rotational transformations\\}

\author{Edward T. Kipreos$^*$ and Riju S. Balachandran$^\dag$}

\address{University of Georgia, 120 Cedar Street, Athens, GA 30602, USA\\
\email{$^*$ekipreos@uga.edu}
\email{$^\dag$balachandranriju@gmail.com}}

\maketitle

\pub{Received 22 December 2020}{Revised 2 April 2021}
\vspace*{-13.5 pt}
\pub{Accepted 14 April 2021}{Published 3 May 2021}

\begin{abstract}
Rotational transformations describe relativistic effects in rotating frames.  There are four major kinematic rotational transformations: the Langevin metric; Post transformation; Franklin transformation; and the rotational form of the absolute Lorentz transformation.  The four transformations exhibit different combinations of relativistic effects and simultaneity frameworks, and generate different predictions for relativistic phenomena.  Here, the predictions of the four rotational transformations are compared with recent optical data that has sufficient resolution to distinguish the transformations.  We show that the rotational absolute Lorentz transformation matches diverse relativistic optical and non-optical rotational data.  These include experimental observations of length contraction, directional time dilation, anisotropic one-way speed of light, isotropic two-way speed of light, and the conventional Sagnac effect.  In contrast, the other three transformations do not match the full range of rotating-frame relativistic observations.

\end{abstract}

\ccode{PACS Nos.: 03.30.+p, 11.30.Cp, 45.40.Bb}
\let\thefootnote\relax\footnote{*Corresponding author.}
\let\thefootnote\relax\footnote{This is an open access article. It is distributed under the terms of the Creative Commons Attribution-NonCommercial-NoDerivs 4.0 (CC BY-NC-ND) License, which permits use and distribution in any medium, provided the original work is properly cited, the use is non-commercial and no modifications or adaptations are made.}

\section{Introduction} \label{section:1}
In 1913, Sagnac demonstrated that within rotating frames, light speed is anisotropic \cite{1}.  In 1914, in an influential response, Witte argued that the light speed anisotropy inherent in the Sagnac effect arose because rotational motion was non-inertial, and the relativity principle (i.e. the Lorentz transformation, LT) was not expected to apply to non-inertial motion \cite{2}.  At the time, Witte's paper implied that rotating frames would lack relativistic effects due to the absolute motion. However, subsequent experiments over the decades have conclusively shown that relativistic effects do occur in rotating frames.  This suggests that the relativistic effects in rotating frames can be described by a rotational transformation (rT).  This brings us to the point of this study: to determine the kinematic rT that accurately describes the relativistic effects that are observed in rotating frames.

Historically, there has been an absence of data of sufficient resolution to distinguish the rTs.  In the absence of discriminating empirical analysis, the Langevin metric has been predominantly used to describe relativity in rotating frames in textbooks \cite{3, 4, 5, 6} and primary publications \cite{7, 8, 9, 10, 11, 12, 13}.  Recent optical data provides the experimental resolutions required to distinguish between the predictions of the rTs.  Our study will assess the Langevin metric and three other rTs to determine which accurately describes relativistic effects in rotating frames.

The four major relativistic kinematic rTs are: the Langevin metric \cite{14, 15}; the Post rT \cite{16}; the ALT rT \cite{17}, which is based on the linear absolute Lorentz transformation (ALT) \cite{18, 19, 20, 21}; and the Franklin rT \cite{22}, which is based on the linear LT \cite{23}.  The four rTs differ in their relativistic effects (length contraction and/or time dilation) and simultaneity frameworks (differential or absolute).  These differences generate distinct predictions for the measurement of light propagation and other relativistic effects.  

In the following subsections, we provide an overview of the four major relativistic rTs.  Throughout, the movement of light around the disk is constrained to a circular path along the rim of the disk, which is often considered to arise from the use of multiple mirrors to guide the light or optical fibers.  The stationary frame is centered on the non-rotating center of rotation.  

\subsection{The Post rT} \label{1.1}
In 1967, Post created an rT \cite{16} that incorporates time dilation, but not length contraction, in an absolute simultaneity framework:
\begin{equation}
dt' = dt\sqrt {1 - \frac{{{\omega ^2}{r^2}}}{{{c^2}}}} ,
\label{eq:7}
\end{equation}
\begin{equation}
d\theta 'r' = d\theta r - \omega rdt,
\label{eq:8}
\end{equation}
\begin{equation}
dr' = dr,
\label{eq:9}
\end{equation}
\begin{equation}
dz' = dz,
\label{eq:10}
\end{equation} 
where $d\theta$ is radians, $\omega$ is angular velocity, $r$ is the radius; unprimed symbols denote stationary-frame coordinates, and primed symbols denote rotating-frame coordinates.

\subsection{The Franklin rT} \label{1.2}
In 1922, Franklin created an rT by expressing the linear LT transformation with peripheral velocity ($\omega r$) in polar coordinates \cite{22}.  The Franklin rT exhibits time dilation, length contraction, and differential simultaneity, similar to the LT. Franklin addressed the issue of superluminal speeds that can arise if the radius is greatly increased by incorporating the relativistic velocity addition formula to make the effective velocity, $v(r)$, a function of the peripheral velocity:
\begin{equation}
v(r) = c\tanh \left( {\frac{{\omega r}}{c}} \right).
\label{eq:11}
\end{equation}
$v(r)$ is then incorporated into the LT equations expressed in polar coordinates:
\begin{equation}
dt' = {{dt - {{v(r)d\theta r} \over {{c^2}}}} \over {\sqrt {1 - {{v{{(r)}^2}} \over {{c^2}}}} }},
\label{eq:12}
\end{equation}
\begin{equation}
d\theta 'r' = \frac{{d\theta r - v(r)dt}}{{\sqrt {1 - \frac{{v{{(r)}^2}}}{{{c^2}}}} }},
\label{eq:13}
\end{equation}
\begin{equation}
dr' = dr,
\label{eq:14}
\end{equation}
\begin{equation}
dz' = dz.
\label{eq:15}
\end{equation} 

\subsection{The ALT rT} \label{1.3}
The ALT rT predicts time dilation, length contraction, and absolute simultaneity \cite{17}, and is obtained using Franklin's approach of directly expressing the linear transformation\cite{18} with peripheral velocity in polar coordinates:
\begin{equation}
dt' = dt\sqrt {1 - \frac{{{\omega ^2}{r^2}}}{{{c^2}}}},
\label{eq:16}
\end{equation}
\begin{equation}
d\theta 'r' = \frac{{d\theta r - \omega rdt}}{{\sqrt {1 - \frac{{{\omega ^2}{r^2}}}{{{c^2}}}} }},
\label{eq:17}
\end{equation}
\begin{equation}
dr' = dr,
\label{eq:18}
\end{equation}
\begin{equation}
dz' = dz.
\label{eq:19}
\end{equation}

\subsection{The Langevin metric} \label{1.4}
In 1921, Langevin created an rT that uses the following polar coordinate transformation equations: \cite{14, 15}
\begin{equation}
t' = t,
\label{eq:1}
\end{equation}
\begin{equation}
\theta ' = \theta - \omega t,
\label{eq:2}
\end{equation}
\begin{equation}
r' = r,
\label{eq:3}
\end{equation}
\begin{equation}
z' = z.
\label{eq:4}
\end{equation}
The unprimed coordinates represent the stationary frame, and the primed coordinates denote the rotating frame but with stationary-frame values for time and distance \cite{7, 24}.  This is different from the other rTs, for which primed coordinates represent rotating-frame values.  Langevin substituted the non-relativistic equations (\ref{eq:1})--(\ref{eq:4}) into the relativistic Minkowski line element expressed in polar coordinates:
\begin{equation}
d{s^2} = {c^2}d{t^2} - d{r^2} - {r^2}d{\theta ^2} - d{z^2}.
\label{eq:5}
\end{equation}
The resulting Langevin metric line element \cite{15} is:
\begin{equation}
ds{'^2} = \left( {{c^2} - r{'^2}{\omega ^2}} \right)dt{'^2} - 2r{'^2}d\theta '\omega dt' - dr{'^2} - r{'^2}d\theta {'^2} - dz{'^2}.
\label{eq:6}
\end{equation}

The Langevin metric exhibits absolute simultaneity as shown by the absence of a distance component in the time Eq. (\ref{eq:1}), which would be required to offset time with distance to generate differential simultaneity, also known as the relativity of simultaneity \cite{25, 26}.  The Langevin metric predicts time dilation and a circumference that is greater than $2 \pi r$ from the rotating-frame perspective \cite{3, 7, 8}.  This is due to the length contraction of rulers in the rotating frame coupled with no change in the circumference length from the stationary perspective, resulting in more length-contracted rulers fitting around the circumference \cite{8}.  From the stationary perspective, while rulers are length contracted on the rim of the rotating disk, the corresponding increase in the number of rulers that can be placed on the rim translates to the absence of net length contraction \cite{8}.  

\subsection{The structure of the rTs and experimental observations} \label{1.5}
The structures of the Franklin rT and ALT rT, and the relations arising from them, are identical to the linear transformations except for the substitution of peripheral velocity and polar coordinates for linear velocity and coordinates.  Consider the shared LT/ALT linear and rotational time dilation relations \cite{18, 23, 27, 28}:
\begin{equation}
dt' = dt\sqrt {1 - \frac{{{v^2}}}{{{c^2}}}} ,
\label{eq:83}
\end{equation}
\begin{equation}
dt' = dt\sqrt {1 - \frac{{{\omega ^2}{r^2}}}{{{c^2}}}} .
\label{eq:84}
\end{equation}
The rotational time dilation relation equation (\ref{eq:84}) accurately describes time dilation in multiple rotational settings, including the half-life of subatomic particles and atomic clocks in motion around the Earth \cite{29, 30, 31, 32}.  This implies that functionality is preserved when relations are described with peripheral velocity and polar terms.

Further evidence for the equivalence of kinematics in linear and rotational frames comes from the generalized Sagnac effect.  It has been experimentally demonstrated that fiber optic light paths with majority linear segments exhibit the Sagnac effect \cite{33, 34}.  The linear light paths contribute to the Sagnac effect in proportion to their length \cite{33, 34}.  The generalized Sagnac effect is independent of the refractive index, and thus would be expected to hold in a vacuum \cite{33, 34}.  The generalized Sagnac equation is: 
\begin{equation}
{\rm{Sagna}}{{\rm{c}}_{{\rm{gen}}}} = \frac{{2l'v}}{{{c^2}}},
\label{eq:35}
\end{equation}
 where $v$ is the velocity, and  $l'$ is the length of the fiber optic light path in the ``moving'' frame \cite{33, 34}.  This is the linear equivalent of the conventional Sagnac equation:
\begin{equation}
{\rm{Sagnac}} = \frac{{2(2\pi r)(\omega r)}}{{{c^2}}}.
\label{eq:34}
\end{equation}

As shown by these relations, and as described in more detail below, kinematic rTs are able to accurately describe rotational relativistic effects to high resolution with exact solutions, such as the isotropic two-way speed of light and the conventional Sagnac effect equation.  The rTs are able to accomplish these accurate descriptions with straightforward descriptions that do not incorporate non-inertial effects.  More complex approaches to describe relativity in rotating frames have been published, but as we describe in Sec. \ref{section:8.1}, these are incompatible with high-resolution optical data.  This is not to suggest that other approaches will not be capable of producing accurate predictions.  Rather, the limited objective of this study is to identify which of the four major kinematic rTs accurately describes rotating-frame relativistic observations.  

The identification of an empirically valid kinematic rT has inherent utility given the ease of their application arising from their straightforward structures.  As a comparable example, consider the Sagnac effect equation, which has a similarly straightforward description with peripheral velocity and polar coordinates.  Yet despite this simple structure, the Sagnac effect equation provides an exact solution that matches experimental data to the highest resolutions of experimental designs \cite{35}.  

The four major rTs are designed to describe spacetime in the absence of curvature.  Nevertheless, the rTs can accurately predict observed relativistic effects in the presence of the Earth's gravitational field.  This implies that Earth-based relativistic observations can be used to test which rT provides valid predictions.

\section{Overview} \label{section:2}
For each of the four rTs, we derive their associated: one-way speeds of light (Sec. \ref{section:3}); two-way speeds of light (Sec. \ref{section:4}); and Sagnac equations (Sec. \ref{section:5}).  In Sec. \ref{section:6}, we compare the predictions with high-resolution optical data.  In Sec. \ref{section:7}, we consider how the rTs match additional relativistic data.  In Sec. \ref{section:8}, we describe how published approaches for the compatibility of the LT and the Sagnac effect are either invalidated by high-resolution optical data or have internal inconsistencies.  In Sec. \ref{section:9}, we present conclusions.

\section{The rT One-Way Speeds of Light} \label{section:3}
In order to determine how the rTs match optical data, we must know the one-way and two-way speeds of light as well as the Sagnac equation associated with each rT.  In this section, we will derive the one-way speed of light directly from the equations for each rT.  We are not aware of published descriptions of the one-way speed of light for the Post rT or Franklin rT.

The rotating-frame one-way speeds of light ($c'$) associated with each rT can be derived directly from their $d\theta' r'$ and $dt'$ transformation equations based on $c' = d\theta' r'/dt'$, where $d \theta r/dt = c$.  The Post rT $c'$ are shown for co-rotating then counter-rotating directions:
\begin{equation}
c{'_{{\rm{Post}}}} = \frac{{d\theta 'r'}}{{dt'}} = \frac{{d\theta r - \omega rdt}}{{dt\sqrt {1 - \frac{{{\omega ^2}{r^2}}}{{{c^2}}}} }} = \frac{{c - \omega r}}{{\sqrt {1 - \frac{{{\omega ^2}{r^2}}}{{{c^2}}}} }};\frac{{c + \omega r}}{{\sqrt {1 - \frac{{{\omega ^2}{r^2}}}{{{c^2}}}} }}.
\label{eq:110}
\end{equation}

	The ALT rT $c'$ matches that previously reported \cite{17}:
\begin{equation}
c{'_{{\rm{ALT}}}} = \frac{{d\theta 'r'}}{{dt'}} = \frac{{\frac{{d\theta r - \omega rdt}}{{\sqrt {1 - \frac{{{\omega ^2}{r^2}}}{{{c^2}}}} }}}}{{dt\sqrt {1 - \frac{{{\omega ^2}{r^2}}}{{{c^2}}}} }} = \frac{c}{{1 + \frac{{\omega r}}{c}}};\frac{c}{{1 - \frac{{\omega r}}{c}}}.
\label{eq:111}
\end{equation}

The Franklin rT $c'$ is isotropic $c$:
\begin{equation}
c{'_{{\rm{LT}}}} = \frac{{d\theta 'r'}}{{dt'}} = \frac{{\frac{{d\theta r - v(r)dt}}{{\sqrt {1 - \frac{{v{{(r)}^2}}}{{{c^2}}}} }}}}{{\frac{{dt - \frac{{v(r)d\theta r}}{{{c^2}}}}}{{\sqrt {1 - \frac{{v{{(r)}^2}}}{{{c^2}}}} }}}} = c;c.
\label{eq:112}
\end{equation}

The Langevin metric $c'$ can be calculated by setting $ds'\textsuperscript{2} = 0$, $dr'\textsuperscript{2} = 0$, and $dz'\textsuperscript{2} = 0$ in the Langevin metric line element (\ref{eq:6}) and then solving for $d\theta 'r'/dt'$.  This gives one-way speeds of light of $c' = c \mp \omega r$.  However, this does not provide the one-way speed of light that would be observed within rotating frames because the derivation uses the ``coordinate'' primed values that have stationary-frame measures of time and distance \cite{7, 24}.  Converting this equation to the values observed in the rotating frame gives $c' = (c \mp \omega r)/(1 - \omega \textsuperscript{2}r\textsuperscript{2}/c\textsuperscript{2})\textsuperscript{0.5}$ \cite{9, 24}.  Thus, the Post rT and the Langevin metric share the same $c'$ values.

\section{The rT Two-Way Speeds of Light} \label{section:4}
We are not aware of the published descriptions of the two-way speed of light for the Post rT, Langevin metric, or Franklin rT.  The rotating-frame two-way speeds of light associated with each rT can be derived from their $d\theta' r'$ and $dt'$ transformation equations by substitution into the basic two-way speed of light equation, with $c' = d\theta' r'/dt'$:
\begin{equation}
c{'_{{\rm{two - way}}}} = \frac{{l' + l'}}{{t{'_{{\rm{co - rot}}}} + t{'_{{\rm{counter - rot}}}}}} = \frac{{l' + l'}}{{\frac{{l'}}{{c{'_{{\rm{co - rot}}}}}} + \frac{{l'}}{{c{'_{{\rm{counter - rot}}}}}}}}.
\label{eq:44}
\end{equation}

The Post rT two-way speed of light is
\begin{equation}
c{'_{{\rm{two - wayPost}}}} = \frac{{2\pi r + 2\pi r}}{{\frac{{2\pi r}}{{\left( {\frac{{d\theta r - \omega rdt}}{{dt\sqrt {1 - \frac{{{\omega ^2}{r^2}}}{{{c^2}}}} }}} \right)}} + \frac{{2\pi r}}{{\left( {\frac{{d\theta r + \omega rdt}}{{dt\sqrt {1 - \frac{{{\omega ^2}{r^2}}}{{{c^2}}}} }}} \right)}}}} = c\sqrt {1 - \frac{{{\omega ^2}{r^2}}}{{{c^2}}}} .
\label{eq:113}
\end{equation}
As a given set of co-rotating and counter-rotating, one-way speeds of light produce a specific two-way speed of light, and the Langevin metric has the same one-way speeds of light as the Post rT, the Langevin metric also generates this two-way speed of light.

	The ALT rT two-way speed of light is isotropic $c$, as previously reported \cite{17}:
\begin{equation}
c{'_{{\rm{two - wayALT}}}} = \frac{{2\pi r + 2\pi r}}{{\frac{{2\pi r}}{{\left( {\frac{{\frac{{d\theta r - \omega rdt}}{{\sqrt {1 - \frac{{{\omega ^2}{r^2}}}{{{c^2}}}} }}}}{{dt\sqrt {1 - \frac{{{\omega ^2}{r^2}}}{{{c^2}}}} }}} \right)}} + \frac{{2\pi r}}{{\left( {\frac{{\frac{{d\theta r + \omega rdt}}{{\sqrt {1 - \frac{{{\omega ^2}{r^2}}}{{{c^2}}}} }}}}{{dt\sqrt {1 - \frac{{{\omega ^2}{r^2}}}{{{c^2}}}} }}} \right)}}}} = c.
\label{eq:114}
\end{equation}

The Franklin rT isotropic two-way speed of light is isotropic $c$:
\begin{equation}
c{'_{{\rm{two - wayLT}}}} = \frac{{2\pi r + 2\pi r}}{{\frac{{2\pi r}}{{\left( {\frac{{\frac{{d\theta r - v(r)dt}}{{\sqrt {1 - \frac{{v{{(r)}^2}}}{{{c^2}}}} }}}}{{\frac{{dt - \frac{{v(r)d\theta r}}{{{c^2}}}}}{{\sqrt {1 - \frac{{v{{(r)}^2}}}{{{c^2}}}} }}}}} \right)}} + \frac{{2\pi r}}{{\left( {\frac{{\frac{{d\theta r + v(r)dt}}{{\sqrt {1 - \frac{{v{{(r)}^2}}}{{{c^2}}}} }}}}{{\frac{{dt + \frac{{v(r)d\theta r}}{{{c^2}}}}}{{\sqrt {1 - \frac{{v{{(r)}^2}}}{{{c^2}}}} }}}}} \right)}}}} = c.
\label{eq:115}
\end{equation}

\section{The rT-Associated Sagnac Equations} \label{section:5}
We are not aware of a published description of the Sagnac equation for the Franklin rT.  The Sagnac equation associated with each rT can be derived from their $d\theta' r'$ and $dt'$ transformation equations by substitution into the basic rotating-frame Sagnac equation, with $c' = d\theta' r'/dt'$:
\begin{equation}
{\rm{Sagna}}{{\rm{c}}_{{\rm{Rotating}}}} = t{'_{{\rm{co - rot}}}} - t{'_{{\rm{counter - rot}}}} = \frac{{l'}}{{c{'_{\rm{co - rot}}}}} - \frac{{l'}}{{c{'_{\rm{counter - rot}}}}}.
\label{eq:99}
\end{equation}

The Post rT generates Sag\textsubscript{1AS} (named for 1 relativistic effect and absolute simultaneity, AS), as previously reported \cite{16}:
\begin{equation}
{\rm{Sa}}{{\rm{g}}_{{\rm{1AS}}}} = \frac{{2\pi r}}{{\left( {\frac{{d\theta r - \omega rdt}}{{dt\sqrt {1 - \frac{{{\omega ^2}{r^2}}}{{{c^2}}}} }}} \right)}} - \frac{{2\pi r}}{{\left( {\frac{{d\theta r + \omega rdt}}{{dt\sqrt {1 - \frac{{{\omega ^2}{r^2}}}{{{c^2}}}} }}} \right)}} = \frac{{4\pi \omega {r^2}}}{{{c^2}\sqrt {1 - \frac{{{\omega ^2}{r^2}}}{{{c^2}}}} }}.
\label{eq:116}
\end{equation}
As a given set of co-rotating and counter-rotating one-way speeds of light produce a specific Sagnac equation, the Langevin metric also generates Sag\textsubscript{1AS}, as previously shown \cite{10, 11}.  A derivation of the Sagnac effect using the Langevin metric as the GR line element has been reported to generate Sag\textsubscript{2AS} \cite{12, 13}.  However, in the derivation, the value of $d \theta' r'$ is substituted with $2\pi r$ rather than the larger circumference that is associated with the Langevin metric \cite{3, 7, 8} (confirmed in one of the papers \cite{13}).  If the larger value had been used, it would have generated Sag\textsubscript{1AS} (not shown).  

	The ALT rT generates Sag\textsubscript{2AS}, as previously reported \cite{36}:
\begin{equation}
{\rm{Sa}}{{\rm{g}}_{{\rm{2AS}}}} = \frac{{2\pi r}}{{\left( {\frac{{\frac{{d\theta r - \omega rdt}}{{\sqrt {1 - \frac{{{\omega ^2}{r^2}}}{{{c^2}}}} }}}}{{dt\sqrt {1 - \frac{{{\omega ^2}{r^2}}}{{{c^2}}}} }}} \right)}} - \frac{{2\pi r}}{{\left( {\frac{{\frac{{d\theta r + \omega rdt}}{{\sqrt {1 - \frac{{{\omega ^2}{r^2}}}{{{c^2}}}} }}}}{{dt\sqrt {1 - \frac{{{\omega ^2}{r^2}}}{{{c^2}}}} }}} \right)}} = \frac{{4\pi \omega {r^2}}}{{{c^2}}}.
\label{eq:117}
\end{equation}

The Franklin rT generates a null Sag\textsubscript{2DS} (2 relativistic effects and differential simultaneity, DS):
\begin{equation}
{\rm{Sa}}{{\rm{g}}_{{\rm{2DS}}}} = \frac{{2\pi r}}{{\left( {\frac{{\frac{{d\theta r - v(r)dt}}{{\sqrt {1 - \frac{{v{{(r)}^2}}}{{{c^2}}}} }}}}{{\frac{{dt - \frac{{v(r)d\theta r}}{{{c^2}}}}}{{\sqrt {1 - \frac{{v{{(r)}^2}}}{{{c^2}}}} }}}}} \right)}} - \frac{{2\pi r}}{{\left( {\frac{{\frac{{d\theta r + v(r)dt}}{{\sqrt {1 - \frac{{v{{(r)}^2}}}{{{c^2}}}} }}}}{{\frac{{dt + \frac{{v(r)d\theta r}}{{{c^2}}}}}{{\sqrt {1 - \frac{{v{{(r)}^2}}}{{{c^2}}}} }}}}} \right)}} = 0.
\label{eq:118}
\end{equation}

\section{Assessing rT Predictions and Optical Data} \label{section:6}
In this section, we will determine the extent to which the rT predictions of light speed in rotating frames match observations.  The Langevin metric, Post rT, and ALT rT predict anisotropic one-way speeds of light, while the Franklin rT predicts an isotropic one-way speed of light.  In 1925, Michelson and Gale showed that light speed is anisotropic on the surface of the rotating Earth \cite{37, 38}.  This result is consistent with the first three rTs, but is not consistent with the Franklin rT.

The Sagnac effect reflects anisotropy in the rotating-frame speed of light.  The Sagnac equation is the sum of anisotropy in the co-rotating and counter-rotating directions.  The null, velocity-invariant Sagnac effect of the Franklin rT is not supported by the observation of the overt Sagnac effect \cite{39}.  The other three rTs generate overt Sag\textsubscript{1AS} and Sag\textsubscript{2AS} Sagnac effects.  The most accurate measurements of the Sagnac effect for light are induced by the Earth's rotation and have experimental resolutions of $\Delta \omega r/\omega r\textsubscript{Earth}$ = $\sim8$ x 10\textsuperscript{-9} \cite{35}.  As described in our companion study, this resolution is not sufficient to distinguish between Sag\textsubscript{1AS} and Sag\textsubscript{2AS}.

The rTs can be distinguished based on their associated two-way speeds of light.  The high resolution of recent optical resonator experiments is sufficient to distinguish the two-way speeds of light of the rTs.  There are eight optical resonator experiments with resolutions of $\Delta c'\textsubscript{two-way}/c$ $\leq$ 1 x 10\textsuperscript{-15} \cite{40, 41, 42, 43, 44, 45, 46, 47}.  As described in our companion study, these resolutions are sufficient to invalidate the anisotropic two-way speed of light of the Langevin metric and Post rT.  Conversely, the ALT rT and the Franklin rT isotropic two-way speed of light is consistent with the data to the resolutions of the optical resonator experiments.

Thus, Sagnac effect data invalidates the prediction of the Franklin rT, and optical resonator data invalidates the predictions of the Langevin metric and the Post rT.  The ALT rT is compatible with both types of data.

\subsection{Optical resonators and length contraction} \label{section:6.1}
Another approach to assess the validity of the rTs is to consider how their associated relativistic effects impact the maintenance of standing optical waves in optical resonators.  From the stationary perspective, the maintenance of standing waves in optical resonators in the direction parallel to motion requires the relativistic effects of length contraction, time dilation, and the relativistic Doppler shift \cite{48}.  These relativistic effects are shared by the LT and ALT \cite{18, 49, 50}.  The Langevin metric and the Post rT lack net length contraction.  Below, the deviation associated with a lack of length contraction is assessed.  

Optical resonators typically utilize orthogonally-arranged optical cavities.  The frequencies of light required to maintain standing optical waves in the two cavities are compared to determine if they change with the rotational motion of the apparatus. The maintenance of a standing wave requires that the cavity length $(l)$ is a multiple of 1/2 the wavelength $(\lambda)$ \cite{51}.  Optical resonators maintain standing waves irrespective of motion.  From the ``stationary'' perspective, the ``moving''-frame light path is longer in the forward direction and shorter in the backward direction.  Therefore, to maintain a standing wave irrespective of motion, the ratio of the wavelength to the length of the cavity when in motion ($\lambda'/l'$) must be the same as the ratio for the stationary non-moving cavity ($\lambda/l$). That is, $\lambda'/l' = \lambda/l$. 

From the ``stationary'' perspective, the ``moving''-frame wavelength is defined by the relativistic Doppler shift equation.  The ``moving''-frame length of the light path in the co-rotating direction is increased by the ratio $c/(c-\omega r)$ and is subject to length contraction to give a length of $cl(1-\omega ^2 r^2/c^2)^{0.5}/(c-\omega r)$.  Thus,  the ratios $\lambda'/l'$ and $\lambda/l$ are equivalent for light propagation in the co-rotating direction:
\begin{equation}
\frac{{\lambda '}}{{l'}} = \frac{{\lambda \sqrt {\frac{{1 + \frac{{\omega r}}{c}}}{{1 - \frac{{\omega r}}{c}}}} }}{{\frac{{cl\sqrt {1 - \frac{{{\omega ^2}{r^2}}}{{{c^2}}}} }}{{c - \omega r}}}} = \frac{\lambda }{l}.
\label{eq:121}
\end{equation}
The ratios of $\lambda'/l'$ and $\lambda/l$ are also equivalent  for the counter-rotating direction:
\begin{equation}
\frac{{\lambda '}}{{l'}} = \frac{{\lambda \sqrt {\frac{{1 - \frac{{\omega r}}{c}}}{{1 + \frac{{\omega r}}{c}}}} }}{{\frac{{cl\sqrt {1 - \frac{{{\omega ^2}{r^2}}}{{{c^2}}}} }}{{c + \omega r}}}} = \frac{\lambda }{l}.
\label{eq:122}
\end{equation}

If length contraction is absent then the ratio $\lambda'/l'$ is not equivalent to $\lambda/l$ for both the co-rotating (\ref{eq:123}) and counter-rotating (\ref{eq:124}) directions: 
\begin{equation}
\frac{{\lambda '}}{{l'}} = \frac{{\lambda \sqrt {\frac{{1 + \frac{{\omega r}}{c}}}{{1 - \frac{{\omega r}}{c}}}} }}{{\frac{{cl}}{{c - \omega r}}}} = \frac{{\lambda \sqrt {1 - \frac{{{\omega ^2}{r^2}}}{{{c^2}}}} }}{l},
\label{eq:123}
\end{equation}
\begin{equation}
\frac{{\lambda '}}{{l'}} = \frac{{\lambda \sqrt {\frac{{1 - \frac{{\omega r}}{c}}}{{1 + \frac{{\omega r}}{c}}}} }}{{\frac{{cl}}{{c + \omega r}}}} = \frac{{\lambda \sqrt {1 - \frac{{{\omega ^2}{r^2}}}{{{c^2}}}} }}{l}.
\label{eq:124}
\end{equation}
Thus, in the absence of length contraction, the standing wave is not maintained by a factor of $(1 - \omega \textsuperscript{2} r\textsuperscript{2}/c\textsuperscript{2})\textsuperscript{0.5}$.  To determine if this deviation would be experimentally observed, we consider an optical resonator experiment that was carried out in Berlin \cite{47}.  Here the constraints on the optical resonator frequency changes were $\Delta f'/f'$ = $\sim$1 x 10\textsuperscript{-18}, where $f'$ is frequency in the rotating frame, and the wavelength was kept constant to maintain a standing wave \cite{47}.  Given the relation $\lambda f = c$, the constraint of $\sim$1 x 10\textsuperscript{-18} also applies to the rotating-frame two-way speed of light ($\Delta c'/c'$), and if the frequency is kept constant, to $\Delta \lambda'/\lambda'$ \cite{47}.  For the peripheral velocity at the latitude of Berlin (282 m/s), the divergence of $1 - (1 - \omega \textsuperscript{2} r\textsuperscript{2}/c\textsuperscript{2})\textsuperscript{0.5}$ is 4.4 x 10\textsuperscript{-13}.  This divergence would have been detected by the resolution of the Berlin experiment, thus invalidating the Langevin metric and the Post rT, both of which lack net length contraction.

\section{Other Relativistic Data that Supports the ALT rT} \label{section:7}
The ALT rT is compatible with a broad range of rotational relativistic data, while the other rTs each do not match all of the rotational data (Table 1, empirically-valid values in bold).  ALT accurately predicts the conventional Sagnac effect equation and the isotropic two-way speed of light.  ALT predicts directional time dilation, with objects experiencing time dilation when in motion relative to a preferred reference frame (PRF) \cite{52}.  Time dilation has been experimentally observed to occur in an absolute and directional manner in rotating frames, as shown by direct comparisons of atomic clocks before and after flights \cite{29, 30, 53}.  In the vicinity of the Earth, time dilation is directional in relation to the non-rotating Earth-centered inertial (ECI) reference frame, which is centered on the core of the Earth \cite{29, 30}.  ALT is only compatible with experimental data if PRFs for ALT are locally associated with gravitational centers \cite{48, 54}, which in the vicinity of the Earth is the ECI.  Thus, directional time dilation in response to motion relative to the ECI is consistent with ALT.  ALT is also compatible with the inferred length contraction that is required to obtain null results in Michelson-Morley-type, Kennedy-Thorndike-type, and optical resonator experiments \cite{28, 48, 55, 56}.  

In contrast, the Franklin rT is incompatible with the observations of the overt Sagnac effect, the anisotropic one-way speed of light in rotating frames, and directional time dilation.  Both the Post rT and the Langevin metric are incompatible with experimental results that depend on net length contraction or the isotropic two-way speed of light.  Additionally, their predicted Sag\textsubscript{1AS} Sagnac effect and one-way speeds of light are invalidated because they imply an anisotropic two-way speed of light that is incompatible with optical resonator experiments (see our companion study).

\begin{table}
\caption{Relativistic attributes of the rotational transformations}
\label{table1}
\vspace*{7 pt}
\centering\begin{tabular}{llllll}
\hline
Relativistic effect &ALT rT &Franklin rT &Post rT &Langevin metric \\
\hline
Length contraction & \boldsymbol{$\sqrt {1 - {{{\omega ^2}{r^2}} \over {{c^2}}}}$} & \boldsymbol{$\sqrt {1 - \frac{{v{{(r)}^2}}}{{{c^2}}}}$} &None & No net effect\\
Time dilation & \boldsymbol{$\sqrt {1 - {{{\omega ^2}{r^2}} \over {{c^2}}}}$} & \boldsymbol{$\sqrt {1 - \frac{{v{{(r)}^2}}}{{{c^2}}}}$} & \boldsymbol{$\sqrt {1 - {{{\omega ^2}{r^2}} \over {{c^2}}}}$} & \boldsymbol{$\sqrt {1 - {{{\omega ^2}{r^2}} \over {{c^2}}}}$} \\
Sagnac effect & \boldsymbol{${{4\pi \omega {r^2}} \over {{c^2}}}$} & 0 & ${{4\pi \omega {r^2}} \over {{c^2}\sqrt {1 - {{{\omega ^2}{r^2}} \over {{c^2}}}} }}$ & ${{4\pi \omega {r^2}} \over {{c^2}\sqrt {1 - {{{\omega ^2}{r^2}} \over {{c^2}}}} }}$ \\
One-way light speeds & \boldsymbol{$\frac{c}{{1 \pm \frac{{\omega r}}{c}}}$} & $c$ & $\frac{{c \mp \omega r}}{{\sqrt {1 - \frac{{{\omega ^2}{r^2}}}{{{c^2}}}} }}$ & $\frac{{c \mp \omega r}}{{\sqrt {1 - \frac{{{\omega ^2}{r^2}}}{{{c^2}}}} }}$ \\
Two-way light speed & \boldsymbol{$c$} & \boldsymbol{$c$} & $c\sqrt {1 - {{{\omega ^2}{r^2}} \over {{c^2}}}} $ & $c\sqrt {1 - {{{\omega ^2}{r^2}} \over {{c^2}}}} $ \\
Directionality & \textbf{Directional} & Reciprocal & \textbf{Directional} & \textbf{Directional} \\
Simultaneity & \textbf{Absolute} & Differential & \textbf{Absolute} & \textbf{Absolute} \\
\hline
\end{tabular}
\vspace*{-4pt}
\end{table}

\section{Approaches for Compatibility of the LT with the Sagnac Effect} \label{section:8}
The Sagnac effect implies that one-way light speed in rotating frames is anisotropic.  If the LT is applied directly to rotating frames, as in the Franklin rT, then the result is isotropic one-way light speed that generates a null Sagnac effect (shown in Secs. \ref{section:3} and \ref{section:5}), which is not observed.  Here we discuss published approaches that seek to explain how the LT or Franklin rT are compatible with the Sagnac effect.  We show that these approaches are either in conflict with recent high-resolution optical data (category 1) or have internal inconsistencies (categories 2 and 3).

\subsection{Application of the LT to rotating frames} \label{section:8.1}
In the first category of approaches, the LT is applied as linear IRFs to rotating frames \cite{57, 58}.  The hypothesis of locality proposes an infinite sequence of linear IRFs that are instantaneously coincident with a point on the rim of a rotating disk; these are referred to as locally co-moving inertial frames (LCIFs).  LCIFs are proposed to display isotropic light speed when measured at infinitesimal distances, below the level of experimental detection \cite{59, 60}.  The integration of LCIFs over a full rotation is used to generate an overt Sagnac effect.  However, these approaches do not generate the conventional Sag\textsubscript{2AS} equation \cite{61, 62, 63, 64, 65}.  The majority of LCIF approaches generate Sag\textsubscript{1AS} \cite{61, 62, 63, 64}, which is associated with the anisotropic two-way speed of light of the Langevin metric and Post rT that is invalidated by optical resonator data.  Approaches to derive the Sagnac effect by utilizing the Langevin metric or Minkowski (LT) spacetime in a GR-based approach in the absence of spacetime curvature also generate the experimentally invalidated Sag\textsubscript{0AS} or Sag\textsubscript{1AS}\cite{66, 67} (and see Sec. \ref{section:5}).

Another approach that uses LCIFs, termed the Zeno paradox, generates an alternate Sagnac equation, Sag\textsubscript{Zeno} \cite{65}.  In this approach, the LCIFs are affected by Coriolis acceleration. Sag\textsubscript{Zeno} is the difference between the light propagation times in the co-rotating and counter-rotating directions: 
\begin{equation}
Sa{g_{\rm{Zeno}}} = \left( {\frac{{2\pi r}}{c}} \right)\sqrt {1 - \frac{{{\omega ^2}{r^2}}}{{2{c^2}}} + \frac{{2\omega r}}{c}} - \left( {\frac{{2\pi r}}{c}} \right)\sqrt {1 - \frac{{{\omega ^2}{r^2}}}{{2{c^2}}} - \frac{{2\omega r}}{c}}.
\label{eq:77}
\end{equation}
The two-way speed of light for Sag\textsubscript{Zeno} is calculated by dividing the combined distance for light propagation in the two directions in the rotating frame ($2 \pi r$ + $2 \pi r$) by the combined times to traverse the circumference in each direction (shown in Eq. \ref{eq:77}). Upon simplification, this gives:
\begin{equation}
c{'_{\rm{two - wayZeno}}} = \frac{{2c}}{{\sqrt {1 - \frac{{{\omega ^2}{r^2}}}{{2{c^2}}} + \frac{{2\omega r}}{c}}  + \sqrt {1 - \frac{{{\omega ^2}{r^2}}}{{2{c^2}}} - \frac{{2\omega r}}{c}} }}.
\label{eq:78}
\end{equation}

	To determine if the two-way speed of light for the Zeno paradox is compatible with optical resonator data, it is compared to the experimental data obtained in Berlin \cite{47}, for which the Earth's rotational peripheral velocity is 282 m/s. At this peripheral velocity, the Zeno paradox gives a deviation of $\Delta c'\textsubscript{two-wayZeno}/c$ = 6.6 x 10\textsuperscript{-13}. The resolution of the optical resonator experiment, $\Delta c'\textsubscript{two-way}/c$ = $\sim$1 x 10\textsuperscript{-18}, would have detected this divergence. Therefore, the Zeno paradox is not compatible with optical resonator data.

\subsection{The time gap}\label{section:8.2}
The second category attributes the Sagnac effect to the time gap discontinuity that occurs when differential simultaneity is applied to a rotating frame \cite{68, 69, 70, 71, 72, 73}.  We will initially explain the principles in the context of a linear spacetime diagram.  A ``stationary'' LT observer views ``moving''-frame time to be shifted over the length of a ``moving'' platform \cite{25, 48, 74}.  The relativistic time offset ($RTO$) describes the time difference between the ``moving''-frame clock times of the platform that are observed by a ``stationary'' observer \cite{48}.  Along with the definition of the $RTO$ as derived by Resnick ($RTO = \mp vl'/c\textsuperscript{2}$) \cite{25}, the $RTO$ can also be defined as $RTO = \mp vl/c\textsuperscript{2}$ if $l$ is defined as the ``at rest'', ``stationary'' length of the ``moving''-frame platform \cite{48}.  Thus, for a ``moving'' platform of length $l' = 1$, the corresponding ``stationary''-frame length is $l = 1$.  Because the number of length units for this definition of $l$ and $l'$ is equivalent, the $RTO$ equation can be expressed with either unit.  Critically, this definition of $l$ makes the length of one circumference unambiguously $2\pi r$ in calculating the $RTO$, as that is the ``at rest'', ``stationary'' length.  Thus, the $RTO$ in a rotating frame is: $\mp 2 \pi \omega r\textsuperscript{2}/c\textsuperscript{2}$.

In a rotating frame, the $RTO$ produces a discontinuity in time when it reaches one circumference that is referred to as the ``time gap'', ``time lag'', or ``synchronization gap'' \cite{68, 69, 70, 71, 72, 73}.  At the discontinuity, a stationary-frame observer would observe that two adjacent points on the disk differ in rotating-frame time by the $RTO$ magnitude, $2 \pi \omega r\textsuperscript{2}/c\textsuperscript{2}$ (see points A and B in Figs. 1(a) and (b)) \cite{68, 69}.  The presence of the time gap currently lacks experimental support, and it has been suggested that internal inconsistencies make it unlikely to be present in actual rotating frames \cite{75, 76, 77}.

\begin{figure}[t]
\centering\includegraphics[width=5.0in]{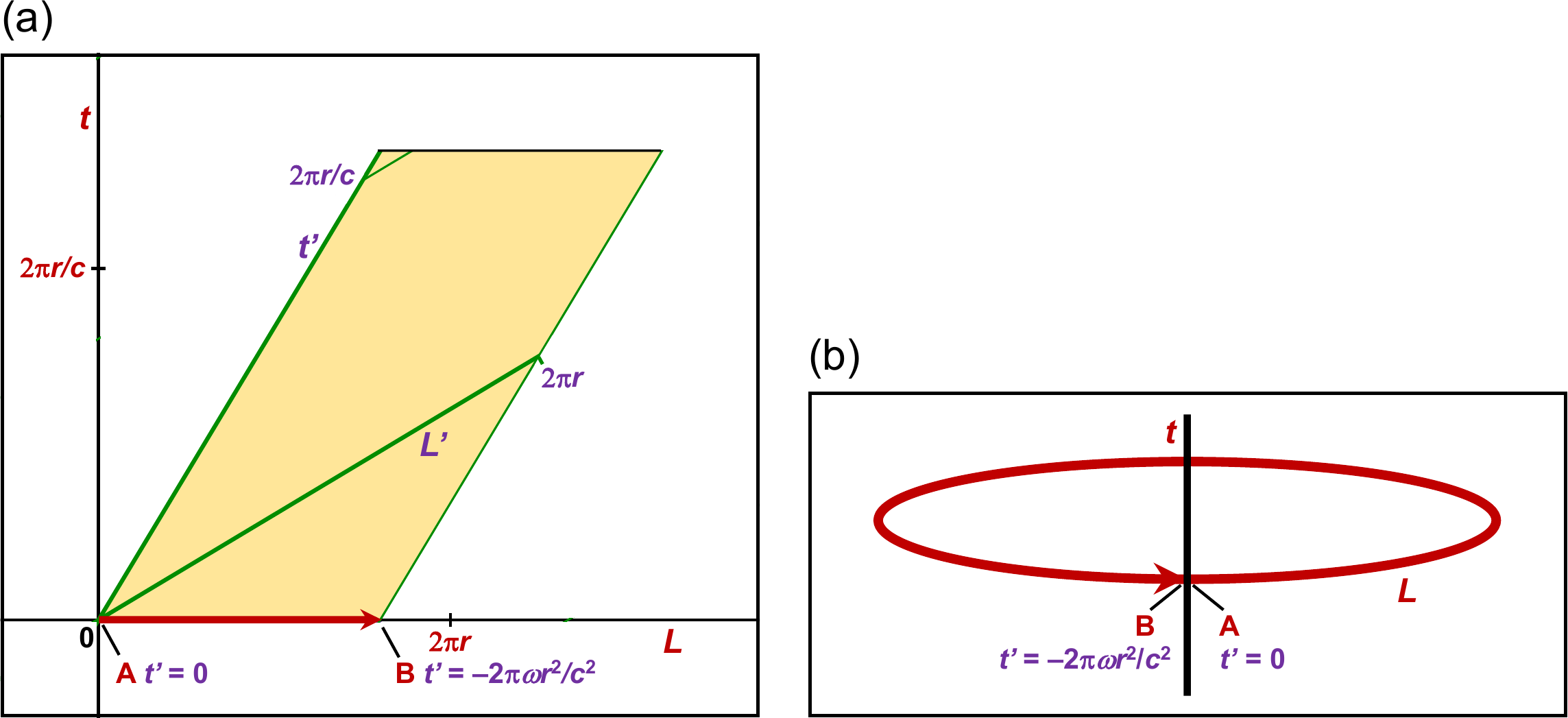}
\caption{The $RTO$ and time gap in rotating frames.  (a) Two-dimensional representation of a Franklin rT rotational spacetime diagram with $\omega r = 0.6c$. Stationary-frame units for distance are in red lettering, and rotating-frame units for distance and times are in purple lettering. The circumference observed by a stationary-frame observer at $t = 0$ is shown as a red arrow.  L denotes arc length.  (b) The $RTO$ represented on a rotating disk, as viewed by a stationary-frame observer at $t = 0$.}
\label{Fig3}
\end{figure}

The $RTO$/time gap has been invoked to explain the Sagnac effect \cite{68, 69, 70, 71, 72, 73}.  These approaches focus on the fact that the magnitude of the $RTO$/time gap is the same magnitude as the unidirectional Sagnac effect (Sagnac\textsubscript{UD}).  However, critically, the $RTO$/time gap and the Sagnac\textsubscript{UD} are of opposite sign.  Anisotropic light speed timing in the co-rotating direction for one circumference is longer than isotropic timing by the Sagnac\textsubscript{UD} value in the co-rotating direction, $+2 \pi \omega r\textsuperscript{2}/c\textsuperscript{2}$.  In contrast, the $RTO$ in the co-rotating direction is $-2 \pi \omega r\textsuperscript{2}/c\textsuperscript{2}$.  

	For light sent across a platform in linear motion, the $RTO$ directly contributes to the isotropic one-way light speed of the LT framework \cite{74}.  From the ``stationary'' perspective, the timing of the light signal in the forward direction is greater than isotropic light speed; but the $RTO$-mediated earlier clock timing of the receiving clock exactly cancels the (otherwise) longer anisotropic timing, allowing the ``moving''-frame observer to calculate isotropic one-way light speed \cite{48, 74}.

If the $RTO$ was present in rotating frames, it would negate the (otherwise) anisotropic timing differences to generate isotropic light speed.  The resulting isotropic light speed would generate a null Sagnac effect.  However, a null Sagnac effect is not experimentally observed \cite{39}.  Conversely, if the $RTO$ was absent from rotating frames then light speed would be anisotropic by the magnitude of the $RTO$ but of opposite sign --- that is because the ability of the $RTO$ to offset clock timing to generate isotropic light speed would be missing.  Notably, experimental observations show that the anisotropy of the one-way light speed in rotating frames is exactly the magnitude of the missing $RTO$ and of opposite sign, i.e. the Sagnac\textsubscript{UD} \cite{35, 78}.  This implies that the $RTO$/time gap is absent from rotating frames.

\subsection{The interconversion of the LT and ALT}\label{section:8.3}
The third category suggests that the Franklin rT and ALT rT are interconvertible alternative forms.  This scenario suggests that converting the Franklin rT to the ALT rT would allow full compatibility with the Sagnac effect \cite{79}.  However, internal inconsistencies occur when differential simultaneity is applied to rotating frames \cite{75, 76, 77}.  This suggests that the Franklin rT and ALT rT describe different physical realities that are not interchangeable because only the absolute simultaneity of ALT is fully compatible with rotating frames \cite{75, 76, 77}.  Further, the interconversion of the two transformations by clock resynchronization is unable to satisfy the expectations of multiple observers \cite{48} or even two coincident observers (see our companion study).

\section{Conclusions}\label{section:9}
Our study assesses the four major kinematic rTs in light of recent high-resolution optical data.  The four rTs differ based on their relativistic effects.  These attributes affect how light propagates in the rotating frame, altering the one-way and two-way speeds of light.  The one-way speeds of light directly translate into specific Sagnac equations.  As shown in our companion study, the relativistic attributes of length contraction, time dilation, and simultaneity that are associated with each rT, when incorporated into basic equations of light propagation, are sufficient to produce the same one-way and two-way speeds of light and Sagnac equation.  This implies that the rT description of these basic attributes of spacetime is sufficient to define the fundamental aspects of light propagation.  

The kinematic rTs have simple structures with descriptions using peripheral velocity and polar coordinates.  Despite this simple structure, the ALT rT produces exact solutions that include the empirically-validated conventional Sagnac effect equation and the isotropic two-way speed of light (see Secs. \ref{section:4} and \ref{section:5}).  Both of these have been confirmed to the resolutions of the experiments.  Additionally, the ALT rT generates a time dilation relation (shared with the Franklin rT) that has also been experimentally confirmed.  Thus, the ALT rT can accurately describe relativistic observations without incorporating non-inertial effects or utilizing co-moving linear transformations.

The Franklin rT incorporates differential simultaneity, which generates an isotropic one-way speed of light in rotating frames and a null Sagnac effect that do not match experimental observations.  In contrast, the Langevin metric and Post rT were designed with the express purpose of generating an overt Sagnac effect \cite{14, 15, 16}, which necessitated the incorporation of absolute simultaneity.  However, these rTs lack net length contraction and generate an anisotropic two-way speed of light that is incompatible with high-resolution optical resonator data.  In contrast, the ALT rT incorporation of absolute simultaneity with both time dilation and length contraction generates an overt Sagnac effect and isotropic two-way speed of light.

The rTs can be distinguished based on their predictions on the ability of rotating-frame observers to determine their rotational motion using light propagation.  The ALT rT predicts that rotating-frame observers can determine their rotational velocity based on anisotropic one-way speeds of light, but cannot determine their rotational velocity using the isotropic two-way speed of light.  The Langevin metric and Post rT predict that rotational velocity can be determined by both anisotropic one-way and anisotropic two-way speeds of light.  The Franklin rT predicts that rotational velocity cannot be determined by either isotropic one-way or isotropic two-way speeds of light.  Significantly, only the ALT rT predictions match observations.  The Earth's rotational velocity can be determined using the anisotropy in the one-way speed of light \cite{35, 37, 38}, but it cannot be determined using the isotropic two-way speed of light \cite{47}.

\end{document}